\documentstyle{article}

\font\tenrm=cmr10
\font\tenit=cmti10
\font\elevenbf=cmbx10 scaled\magstep 1
\font\elevenrm=cmr10 scaled\magstep 1
\font\elevenit=cmti10 scaled\magstep 1

\font\ninerm=cmr9

\renewenvironment{thebibliography}[1]
 { \elevenrm
   \begin{list}{\arabic{enumi}.}
    {\usecounter{enumi} \setlength{\parsep}{0pt}
     \setlength{\itemsep}{3pt} \settowidth{\labelwidth}{#1.}
     \sloppy
    }}{\end{list}}

\begin{document}
\begin{center}
\vglue 0.6cm
{
 {\elevenbf        \vglue 10pt
               REVIEW OF Z' CONSTRAINTS\footnote{
\ninerm
\baselineskip=11pt
Invited talk presented by F. del Aguila at the Workshop on Physics and
Experiments with Linear e$^+$e$^-$ Colliders, Hawaii, April 1993}
\\}
\vglue 5pt

\vglue 0.6cm
{\tenrm F. DEL AGUILA \\}
\baselineskip=13pt
{\tenit Departamento de F\'\i sica Te\'orica y del Cosmos,
Universidad de Granada \\}
\baselineskip=12pt
{\tenit Granada, 18071, Spain \\}

\vglue 0.3cm
{\tenrm M. CVETI\v C and P. LANGACKER\\}
\baselineskip=13pt
{\tenit Department of Physics, University of Pennsylvania \\}
\baselineskip=12pt
{\tenit Philadelphia, PA 19104-6396 \\}
\baselineskip=14pt
{\tenit September 3, 1993, UPR-0583T\\}}

\end{center}

\vglue 0.6cm
{\elevenbf\noindent 1. Introduction}
\vglue 0.4cm
\elevenrm
No signal of a new gauge boson has been observed up to now. Thus, although
many extensions of the standard model require new gauge bosons, only limits
on their interactions exist. We review {\elevenit present Z' bounds}
(Section 2), discuss {\elevenit future Z' constraints} (Section 3) from
TEVATRON, HERA and LEP, and give a brief review of the {\elevenit Z'
diagnostics at future colliders} (Section 4).

\vglue 0.6cm
{\elevenbf\noindent 2. Present Z' Bounds}
\vglue 0.4cm
Grand unified theories (GUT's) based, {\elevenit e.g.}, on the gauge group
{\elevenit SO(10)} or {\elevenit E}$_6$, as well as those based on superstring
theory predict new gauge bosons $^1$. However, at present there are only
(very model dependent) limits on their interactions $^2$.
We review present constraints on new {\elevenit Z'},
presenting numerical results for the typical GUT, superstring-motivated,
and left-right symmetric
models: $\chi ,\psi ,\eta , LR$ (their definition
is given for instance in Ref. 3). In Table 1 we collect the 95$\% $
{\elevenit C.L.} direct bounds on $M_{Z'}$ from TEVATRON,
corresponding to an integrated luminosity of $\sim 4 \ pb^{-1}$ $^4$.
Indirect bounds are given in columns two and three. These
update the global fits of Ref. 5
by including the 1992 LEP data $^6$ and other new results.
(An update of the fits in Ref. 7
which use a larger data set yields similar results to those
presented here.) The different
data sets contributing to the total $\chi ^2$ are gathered in Table 2. The
unconstrained (constrained) bounds do not (do) assume a definite, minimal
Higgs sector, and thus an unconstrained (constrained) {\elevenit Z'Z}$^0$
mixing angle, $s_3$. In both cases the top and the Higgs masses and the
strong coupling constant are free parameters.

\rightskip=3pc
\leftskip=3pc
{\tenrm\baselineskip=12pt
\noindent
Table 1. Present bounds on the mass of new gauge bosons (in {\tenit GeV})
at $ 95 \% $ {\elevenit C.L.}. }

\rightskip=0pc
\leftskip=0pc
\vglue 0.1cm
\begin{center}
\begin{tabular}{|c|ccc|}
\hline
\multicolumn{1}{|c}{} & direct
& $\begin{array}{c} {\rm indirect} \\ {\rm (unconstrained)} \end{array}$ &
\multicolumn{1}{c|}{$\begin{array}{c} {\rm indirect} \\ {\rm (constrained)}
\end{array}$} \\
\hline
$\chi$ & $320$ & $380$ &
$670$ \\
$\psi$ & $300$ & $200$ &
$200$ \\
$\eta$ & $310$ & $210$ &
$440$ \\
$LR$ & $350$ & $430$ &
$990$ \\
\hline
\end{tabular}
\end{center}
\vglue 0.1cm

\rightskip=3pc
\leftskip=3pc
{\tenrm\baselineskip=12pt
\noindent
Table 2. Data sets used to obtain the indirect $M_{Z'}$ limits.
Neutral-current parameters, $\nu q, \nu _{\mu }e, eH$, and gauge boson
masses are summarized in Ref. 2. LEP data were presented
at the XXVIIIth Rencontre de Moriond $^6$}.

\rightskip=0pc
\leftskip=0pc
\vglue 0.1cm
\begin{center}
\begin{tabular}{|c|ccc|}
\hline
\multicolumn{1}{|c}{} & Quantity &
Experimental Value &
\multicolumn{1}{c|}{Correlation Matrix} \\
\hline
$\nu q$ & $\begin{array}{c} g_L^2 \\ g_R^2 \\
\theta _L \\ \theta _R \end{array}$ &
$\begin{array}{c} 0.3003 \pm 0.0039 \\ 0.0323 \pm 0.0033 \\
2.49 \pm 0.037 \\ 4.69 \pm 0.38 \end{array}$ &
$\begin{array}{cccc} 1. &  &  &  \\  & 1. &  &  \\
 &  & 1. &  \\  &  &  & 1. \end{array}$ \\
\hline
$\nu _{\mu} e$ & $\begin{array}{c} g_A^e \\ g_V^e \end{array}$ &
$\begin{array}{c} -0.508 \pm 0.015 \\ -0.035 \pm 0.017 \end{array}$ &
$\begin{array}{cc} 1. & -0.04 \\ -0.04 & 1. \end{array} $ \\
\hline
$eH$ & $\begin{array}{c} C_{1u} \\ C_{1d} \\
C_{2u} - \frac{1}{2}C_{2d} \end{array}$ &
$\begin{array}{c} -0.214 \pm 0.046 \\ 0.359 \pm 0.041 \\
-0.04  \pm 0.13   \end{array}$ &
$\begin{array}{ccc} 1. & -0.995 & -0.79 \\ -0.995 & 1. & 0.79 \\
-0.79 & 0.79 & 1. \end{array}$ \\
\hline
$p\bar p$ & $\begin{array}{c} M_W \\ M_W/M_Z \end{array}$ &
$\begin{array}{c} 79.91 \pm 0.39 \ GeV \\ 0.8813 \pm 0.0041 \end{array}$ &
$\begin{array}{cc}  &  \\  &  \end{array}$ \\
\hline
LEP & $\begin{array}{c} M_Z \\ \Gamma _Z \\ \sigma ^0_h \\
R_l \\ A_{FB} \end{array}$ &
$\begin{array}{c} 91.187 \pm 0.007 \ GeV \\ 2488 \pm 7 \ MeV \\
41.446 \pm 0.169 \ nb \\ 20.833 \pm 0.056 \\ 0.016 \pm 0.002 \end{array}$ &
$\begin{array}{ccccc} 1. & -0.154 & 0.023 & 0.012 & 0.070 \\
-0.154 & 1. & -0.143 & 0.007 & 0.005 \\ 0.023 & -0.143 & 1. & 0.126 &
0.003 \\ 0.012 & 0.007 & 0.126 & 1. & 0.008 \\ 0.070 & 0.005 & 0.003 &
0.008 & 1. \end{array}$ \\
\hline
%
%
\end{tabular}
\end{center}
\vglue 0.1cm

Indirect $M_{Z'}$ limits have a large, model dependent correlation with
the other variables in the fit.
However, the value of the weak angle and upper limits on the top
quark mass are insensitive to large variations of the $M_{Z'}$
values. In general LEP data put stringent limits on the {\elevenit Z'Z}$^0$
mixing angle $^{5,7,8}$.

Many extensions of the standard model, like the ones considered here,
predict the existence of new fermions. The {\elevenit Z'} limits, however,
change little when
fermion mixing between standard and new (vector-like) fermions is
allowed $^9$.

Stringent limits on new interactions can also be derived from astrophysical
constraints. The corresponding limits on the {\elevenit Z'} mass are of the
order of 2 - 3 {\elevenit TeV} $^{10}$, but they rely on the existence
of an almost massless right-handed neutrino and can be avoided.
Baryogenesis may eventually  also pose a problem for extended gauge theories
with heavy neutrinos and new gauge bosons in the {\elevenit TeV} range
$^{11}$.

\vglue 0.6cm
{\elevenbf\noindent 3. Future Z' Constraints}
\vglue 0.4cm
In Table 3 we present the estimates for the forthcoming bounds on $M_{Z'}$.
The limits for TEVATRON correspond to an integrated luminosity, already
accumulated, $\int {\cal L} dt = 25 \ (100) \ pb^{-1}$; whereas the bounds
at HERA assume polarized beams and an integrated luminosity
$\int {\cal L} dt = 100 \ pb^{-1}$.

\vglue 0.2cm
{\elevenit \noindent 3.1. TEVATRON}
\vglue 0.1cm
The direct bounds on $M_{Z'}$ are robust $^{7,12}$. They do not depend on
the {\elevenit Z'Z}$^0$ mixing angle and are somewhat insensitive
to the top and Higgs masses.

\newpage
\rightskip=3pc
\leftskip=3pc
{\tenrm\baselineskip=12pt
\noindent
Table 3. Future constraints on the mass of new gauge bosons (in {\tenit GeV}).
The blank in the 95\%  {\tenit C.L.} HERA bounds indicates no sensitivity
to {\tenit Z'} masses above the {\tenit Z} mass}.

\rightskip=0pc
\leftskip=0pc
\vglue 0.1cm
\begin{center}
\begin{tabular}{|c|cc|}
\hline
\multicolumn{1}{|c}{}
& $\begin{array}{c} {\rm TEVATRON} \\  \int {\cal L} dt = 25 \ \ \
(100) \ pb^{-1} \ \ \ \end{array}$ & \multicolumn{1}{c|}{HERA} \\
\hline
$\chi$ & $470 \ \ \ (620)$ & $240$ \\
$\psi$ & $450 \ \ \ (600)$ & $-$ \\
$\eta$ & $460 \ \ \ (610)$ & $180$ \\
$LR$ & $510 \ \ \ (660)$ & $370$ \\
\hline
\end{tabular}
\end{center}
\vglue 0.1cm

\vglue 0.2cm
{\elevenit \noindent 3.2. HERA}
\vglue 0.1cm
The sensitivity of HERA to the exchange of a heavy neutral
{\elevenit Z'} boson and its discovery potential has been
systematically analysed in Ref. 13. The corresponding limits
on new gauge interactions are not competitive, for
instance, with those from TEVATRON (see Table 3). The diagrams
involved are the same in both cases but new gauge bosons are
produced in the {\elevenit s}-channel at hadron colliders and
exchanged in the {\elevenit t}-channel at lepton-hadron
colliders. In contrast with TEVATRON, HERA is sensitive to
the relative sign of {\elevenit Z'} couplings. Due to the
different weight of gauge couplings, HERA also enhances its
relative potential for definite choices of gauge couplings.
At any rate the limits expected at TEVATRON rule out the
possibility of observing a {\elevenit Z'} at HERA first.

\vglue 0.2cm
{\elevenit \noindent 3.3. LEP}
\vglue 0.1cm
A similar fit to the one in Table 1 but with improved LEP
errors makes even more apparent the comments in the previous
Section. For the unconstrained case
a global fit to precise standard model data gives only a weak
{\elevenit Z'} mass limit.

\vglue 0.6cm
{\elevenbf\noindent 4. Z' Diagnostics at Future Colliders}
\vglue 0.4cm
Large colliders can probe new gauge interactions for {\elevenit Z'}
masses up to several {\elevenit TeV}. If the samples are large enough
the determination of {\elevenit Z'} gauge couplings to matter can be
attempted. The hadron and lepton colliders are discussed separately
below for the case of one extra neutral gauge boson coupled minimally.
(Large colliders will also allow for measuring non-minimal
coefficients of a general parametrization with effective operators
(see for instance Ref. 14).)

\vglue 0.2cm
{\elevenit \noindent 4.1. LHC/SSC}
\vglue 0.1cm
A detailed discussion is presented by M. Cveti\v c in this Workshop
$^{15}$. {\elevenit Z'} physics is described by eight parameters:
the {\elevenit Z'} mass and width, $M_{Z'}, \Gamma _{Z'}$, five
gauge couplings, $g_{Z'}, \ \gamma ^l_L \equiv \frac {(g^l_{Z'L})^2}
{(g^l_{Z'L})^2+(g^l_{Z'R})^2}, \ \gamma ^q_L \equiv \frac {(g^q_{Z'L})^2}
{(g^l_{Z'L})^2+(g^l_{Z'R})^2}, \ \tilde U \equiv (\frac{g^u_{Z'R}}
{g^q_{Z'L}})^2, \ \tilde D \equiv (\frac{g^d_{Z'R}}
{g^q_{Z'L}})^2$ (no sensitivity is expected at LHC/SSC to
the sign of the {\elevenit Z'} gauge couplings to quarks
and leptons), and the {\elevenit Z'Z}$^0$ mixing angle, $s_3$.
The cross section for $pp \rightarrow Z' \rightarrow l^+l^-$ determines
the $Z'$ mass, width, and gauge coupling. Combining the ratio of this
cross section in two rapidity bins with the forward-backward asymmetry,
the rare decay modes $Z' \rightarrow Wl\nu _l$, and three associated
productions $pp \rightarrow Z'V (V=Z,W,\gamma )$, and assuming
inter-family universality, small {\elevenit Z'Z}$^0$ mixing, and the
{\elevenit Z'} charge commuting with the {\elevenit SU(2)}$_L$ generators,
three out of four normalized couplings could be extracted. $\gamma ^q_L$
requires the measurement of the $pp \rightarrow Z' \rightarrow q\bar q$
cross section. This is a difficult task; however, it has been recently
claimed that it may be possible $^{16}$. Finally, with appropriate cuts, $s_3$
may be measured by studying the rare $Z'$ decays into two charged leptons
plus two neutrinos $^{17}$.
Hence, except for the signs of the gauge couplings to quarks and leptons,
all parameters fixing the interactions of a new neutral gauge boson with a
mass $\sim 1-2\ TeV$ may be determined at large hadron colliders.

\vglue 0.2cm
{\elevenit \noindent 4.2. e$^+$e$^-$}
\vglue 0.1cm
LEP 200 will measure $M_W$ with high precision, improving the top mass and
eventually the Higgs mass limits and the indirect $M_{Z'}$ lower bounds.
Larger {\elevenit e$^+$e$^-$} colliders may produce a new {\elevenit Z'}
on-shell, but a more realistic scenario may be the production of a new
heavy gauge boson far off-shell. (A detailed discussion of this case can
be found in Ref. 18 (see also Ref. 19).) In both cases the
$e^+e^- \rightarrow Z' \rightarrow W^+W^-$ channel
offers the possibility of measuring the {\elevenit Z'Z}$^0$
mixing angle, $s_3 \ ^{20}$. However, the $s_3$ bounds seem to exclude this
possibility for an {\elevenit e$^+$e$^-$} collider with a center of mass
energy of 500 {\elevenit GeV} (NLC).
On the other hand, it is claimed that
the $e^+e^- \rightarrow Z' \rightarrow f\bar f$ channel distinguishes
between extended gauge models for {\elevenit Z'} masses up to 3
{\elevenit TeV} at NLC $^{18}$. This requires taking into account radiative
corrections and a good control of the experimental set up. Similarly to the
LHC/SSC case  one can determine (some of) the parameters describing the
{\elevenit Z'} interactions. For a {\elevenit Z'} much heavier than 500
{\elevenit GeV} ($\sim $ 1 {\elevenit TeV})
no determination of the {\elevenit Z'} width and mass
seem to be possible, since the {\elevenit Z'} amplitudes are proportional to
$\frac{g^2_{Z'}}{M^2_{Z'}}$. However, with final state quark
identification $\sigma ^l, \ R = \frac
{\sigma ^h}{\sigma ^l}$,
and $A ^{l,h}_{FB}$ will allow one to measure not only
$\frac{g^2_{Z'}}{M^2_{Z'}}$ and $\gamma ^q_L , \ \gamma ^l_L, \ \tilde U ,
\ \tilde D$, but will also provide a unique determination of the relative signs
of the hadronic and leptonic couplings $^{21}$.

\vglue 0.6cm
{\elevenbf \noindent 5. Acknowledgements \hfil}
\vglue 0.4cm

We thank F. Cornet, J. Liu,
M. Mart\'\i nez, J. Moreno and M. Quir\'os for discussions.
\vglue 0.5cm

{\elevenbf\noindent 6. References \hfil}
\vglue 0.4cm

\end{document}